\documentclass[aps,twocolumn,showpacs,superscriptaddress,amsmath]{revtex4}
\usepackage{amsmath,amsfonts,euscript,amssymb}

\newcommand{\be}{\begin{equation}}
\newcommand{\ee}{\end{equation}}
\newcommand{\bea}{\begin{eqnarray}}
\newcommand{\eea}{\end{eqnarray}}

\begin{document}
\title{
 A Condensate Mechanism of Conformal Symmetry Breaking and Higgs Particle
}

\author{V.N.~Pervushin}
\affiliation{Bogoliubov Laboratory of Theoretical Physics, Joint Institute for
Nuclear Research, 141980 Dubna, Russia}
\author{A.B.~Arbuzov}
\affiliation{Bogoliubov Laboratory of Theoretical Physics, Joint Institute
for Nuclear Research, 141980 Dubna, Russia}
\author{R.G.~Nazmitdinov}
\affiliation{Department de F{\'\i}sica,
Universitat de les Illes Balears, E-07122 Palma de Mallorca, Spain}
\affiliation{Bogoliubov Laboratory of Theoretical Physics, Joint Institute
for Nuclear Research, 141980 Dubna, Russia}
\author{A.E.~Pavlov}
\affiliation{Bogoliubov Laboratory of Theoretical Physics, Joint Institute
for Nuclear Research, 141980 Dubna, Russia}
\affiliation{Moscow State Agri-Engineering University, 127550 Moscow, Russia}
\author{A.F.~Zakharov}
\affiliation{Bogoliubov Laboratory of Theoretical Physics, Joint Institute
for Nuclear Research, 141980 Dubna, Russia}
\affiliation{ Institute of Theoretical and Experimental Physics, 117259 Moscow, Russia}

\begin{abstract}
A mechanism of spontaneous conformal symmetry breaking based on field condensates
in the Standard Model of strong and electroweak interactions is suggested.
It is shown that an existence of the top quark condensate can supersede the
tachyon mass term in the Higgs potential in the standard mechanism of
electroweak symmetry breaking. 
Considering the ratio of a field condensate to the corresponding mass power
(depending on quantum statistics) for various fields as a conformal invariant, we obtain
the Higgs boson mass to be about $130\pm 15$ GeV.
\end{abstract}

\date{\today}
\maketitle

Recently a few research groups reported upon the discovery of scalar particles
with almost similar masses around $120-140$ GeV \cite{:2012gk,:2012gu,Aaltonen:2012qt}.
The experimentalists use an extreme caution in
the identification of these particles with the long-waiting Higgs one.
Indeed, the literature contains a plethora of predictions on
lower and upper limits of the Higgs mass based
on many different ideas, models and numerical techniques, which
are close to the observed values.
The question on a genuine mechanism which provides
an unambiguous answer about the Higgs mass
is a real challenge to high energy physics and is crucial
for the base of the Standard Model (SM) \cite{SM}.
It is  especially noteworthy that in the SM
the Higgs mass is introduced {\it ad hoc}.

According to a general wisdom, all SM particles (may be except neutrinos) own
masses due to the spontaneous symmetry breaking (SSB)
of the electroweak gauge symmetry~\cite{higgs}.
In particular, one deals with the potential (in notation of Ref.~\cite{Beringer:1900zz}):
 \be
 \label{5a}
 V_{\rm Higgs}(\phi)=\frac{\lambda^2}{2}(\phi^\dagger\phi)^2 + \mu^2\phi^\dagger\phi,
 \ee
where one component of the complex scalar doublet field
$\phi=\left(\begin{array}{c}\phi^+\\ \phi^0\end{array}\right)$
acquires a non-zero vacuum expectation value $\langle\phi^0 \rangle = v/\sqrt{2}$
if $\mu^2<0$ (the stability condition $\lambda^2>0$ is always assumed).
Note that the tachyon-like mass term in the potential is critical for this construction.
In contrast to the SSB, it breaks the conformal symmetry explicitly being
the only {\em fundamental} dimension-full parameter in the SM.
We recall that the explicit conformal symmetry breaking in the
Higgs sector gives rise to the unsolved problem of fine tuning in the renormalization
of the Higgs boson mass.
That is certainly one of the most unpleasant features of the SM.

In the classical approximation, from the condition of the potential minimum
one obtains the relation between the vacuum expectation value and the primary parameters
$\mu$ and $\lambda$ in the form $v=\sqrt{-2{\mu^2}}\,/\lambda$.
This quantity can be defined as well with the aid of the Fermi coupling constant
derived from the muon life time measurements:
$v=(\sqrt{2}G_{\mathrm{Fermi}})^{-1/2}\approx 246.22$~GeV.
The experimental studies at LHC~\cite{:2012gk,:2012gu}
and Tevatron~\cite{Aaltonen:2012qt}
observe an excess of events in the data compared with the background
in the mass range around $\sim 126$ GeV.
Taking into account radiative corrections, such a mass value
makes the SM being stable up to the Planck mass energy scale~\cite{Bezrukov:2012sa}.
Nevertheless, the status of the SM and the problem of
the mechanism of elementary particle mass generation are still unclear.

The idea on dynamical breaking of the electroweak gauge symmetry
with the aid of the top quark condensate was continuously discussed in the literature since
the pioneering papers~\cite{Nambu:1989sx,Nambu:1990hj,Bardeen:1989ds}
(see also review~\cite{Cvetic:1997eb}, 
recent papers~\cite{Volovik:2012qq,Matsuzaki:2012mk},  and references therein).
Such approaches suffer, however, from formal quadratic divergences
in tadpole loop diagrams leading, in particular, to the
naturalness problem (or fine tuning) in the renormalization of the Higgs
boson mass.

All mentioned facts suggest that, it might be well to examine that the SSB
is also responsible for the Higgs field energy scale.
To begin with, we suppose that there
is a general mechanism of the SSB, which is responsible for the appearance
of all SM field condensates.
The main feature of our approach is the assumption about the underlying
(softly broken) conformal symmetry which protects
the jump of the Higgs boson mass to a cut-off scale.
We will call this mechanism the spontaneous conformal symmetry breaking (SCSB).
Evidently, in this case one should require the conservation
of the conformal symmetry of the genuine theory fundamental Lagrangian.
It will be shown that the SCSB provides the breaking of the gauge, chiral,
and conformal symmetries on equal footing.
Therefore, it allows also to introduce the universal relation
between different condensates determined relative to the corresponding mass power
depending on quantum statistics, see Eq.~(\ref{13}). Our basic conjecture is
that this relation holds approximately after the spontaneous breaking of the
conformal symmetry.

Following the ideas of Nambu \cite{Nambu:1989sx,Nambu:1990hj}, we
generate the SCSB of the Higgs potential, using the top quark condensate.
It is assumed that the general construction of the SM
should remain unchanged. Let us start with the conformal invariant Lagrangian
of Higgs boson interactions
 \be \label{L_int}
 L_{\mathrm{int}} = - \frac{\lambda^2}{8}h^4 - g_t h~\bar{t}t.
 \ee
Here, for the beginning, we consider only the most intensive
terms: the self-interaction and the Yukawa ones of the top quark coupling
constant $g_t$. 
Contributions of other interaction terms will be considered below as well.
We assume that the $O(4)$ symmetry of the Higgs
sector has being spontaneously broken to the $O(3)$ symmetry.
So that the whole construction should contain the appereance of the non-zero
Higgs field vacuum expectation value.

In accord with the postulates of Quantum Field Theory (QFT),
to calculate physical quantities (including the mass spectrum)
one has to apply the normal ordering to field operators.
The normal ordering  in the Hamiltonian of interacting scalar fields leads
to the condensate density ${\langle \phi \phi \rangle}_{\rm Cas}$
\be\label{C-1}
{\langle \phi \phi \rangle}_{\rm Cas}
=\!\frac{1}{V_0}
 \sum_p\frac{1}{2\sqrt{p^2+m^2}}\,,
\ee
which we name the Casimir condensate density.
Indeed, by means of differentiation
\be\label{C-2}
{\langle \phi \phi \rangle}_{\rm Cas}=\frac{2}{V_0}\dfrac{\partial }{\partial m^2}E_{\rm Cas}\,,
\ee
this density is related to the Casimir energy \cite{Bordag}
\be
E_{\rm Cas}=\frac{1}{2}\sum_{\bf p} \sqrt{{p}^2+m^2}.
\ee
In the continual limit of the QFT one has
\bea\label{universality}
&&\frac{1}{V_0}\sum_{\bf p}\frac{1}{2\sqrt{p^2+m_t^2}}\Rightarrow
\int \frac{d^3p}{(2\pi)^3}\frac{1}{2\sqrt{p^2+m^2}}=\nonumber\\
&&=m^2\int \frac{d^3x}{(2\pi)^3}\frac{1}{2\sqrt{x^2+1}}\equiv\gamma_0 \cdot m^{2}.
\eea
Thus, the Casimir condensate density
of a massive scalar field in the absence of any additional scale
is proportional to its squared mass
\be\label{Cas-2}
\langle \phi \phi\rangle_{\rm Cas}=\gamma_0 \cdot m^{2} \Rightarrow
\frac{\langle \phi \phi\rangle_{\rm Cas}}{m^{2}}\equiv \gamma_0\, ,
\ee
where $\gamma_0$ is {\it a dimensionless conformal quantity}
with a zero conformal weight (see discussion on conformal weights in \cite{grg}).

The normal ordering of a fermion pair (we intentionally interchange the order of
fermion fields to deal with positive condensates)
$f \bar f=:f \bar f:+ \langle f \bar f \rangle$
yields the condensate density of the fermion field $\langle f\bar f \rangle$
in the  Yukawa interaction term in Eq.(\ref{L_int}).
In virtue of the above results, we have
for the top quark Casimir condensate density
 \be\label{11}
 {\langle t\bar t \rangle}_{\rm Cas}= 4 N_c \frac{m_t}{V_0}\sum_{\bf p}
 \frac{1}{2\sqrt{p^2+m_t^2}}=4N_c\cdot \gamma_0 \cdot m_t^{3},
 \ee
where $N_c=3$ is the number of colors.

Keeping in mind all these results, we are ready to treat the contribution
of  the top quarks to the effective potential, generated by the term  Eq.(\ref{L_int}):
 \bea\label{V_cond}
 V_{\rm cond}(h) &=& \frac{\lambda^2}{8}h^4 -g_t\langle t\bar t\,\rangle h.
 \eea
The extremum condition for the potential $dV_{\rm cond}/dh|_{h=v}=0$
yields the relation
 \be \label{lambda}
 {v^3}\frac{\lambda^2}{2} = {g_t\langle t\bar t\,\rangle}.
 \ee
This relation follows from the fact that
the Higgs field has a zero harmonic $v$ in the standard decomposition
of the field $h$ over harmonics  $h = v  + H$, where $H$ is the sum of all nonzero
 harmonics with a condition $\int d^3x H=0$. Here, the Yukawa coupling of
the top quark $g_t \approx 1/\sqrt{2}$ is known from the experimental value
of top quark mass $m_t=v g_t\simeq 173.4$~GeV.

The spontaneous symmetry breaking yields
the potential minimum which results in the nonzero
vacuum expectation value $v$ and Higgs boson mass.
The substitution $h = v  + H$ into the potential~(\ref{V_cond})
leads to the result
\be
V_{\rm cond}(h)\!=\!V_{\rm cond}(v)+\frac{m_H^2}{2}H^2+
\frac{\lambda^2 v}{2} H^3+\frac{\lambda^2}{8}H^4,
\ee
which defines the scalar particle mass as
\be
\label{mh}
m_H^2\equiv\frac{\lambda^2}{2}3v^2 .
\ee
 We stress that this relation is different from
the one $(m_H=\lambda v)$ which emerges in the SM with the Higgs potential~(\ref{5a}).

With the aid of Eqs.(\ref{lambda}),(\ref{mh}),
 the squared scalar particle mass can be expressed in terms of the $t$
quark condensate:
\be\label{h_mass}
m^2_H = \frac{3 g_t \langle t\bar t\,\rangle}{v}.
\ee
The conjecture on universality of the conformal invariant ratios of the field condensate densities and the
corresponding mass powers (see Eqs. (\ref{universality})--(\ref{11})),
allows us to determine the $t$ quark condensate density with the aid of the light quark one.
Using the relation
\be
\label{13}
\frac{\langle t\bar t \rangle}{m^3_{t }}=\frac{\langle q\bar q \rangle}{m^3_{q} }
 \ee
we consider the left and right hand sides as scale invariants. However their
numerators and denominators are scale dependent. Therefore we have to choose the proper scales.
For the left hand side, the scale is naturally defined by the known $t$ quark mass. We define
the scale of the right hand side by
the light quark condensate  density $\langle q\bar q \rangle$. It is rather
accurately determined in the chiral limit of
the QCD low-energy phenomenology \cite{Beringer:1900zz}:
\be
\label{13c}
\langle q \bar{q} \rangle\simeq (250~{\rm MeV})^3.
\ee
At this scale the light quark possesses the constituent mass $m_{q}\approx 330$~MeV
estimated in the QCD-inspired model~\cite{gmor1}.
With the aid of
Eq.(\ref{13}) one determines the top quark condensate value
\be
\langle t\bar t\,\rangle \approx (126\ {\mathrm{GeV}})^3.
\ee
Such a large value of the top quark condensate does not affect the
low energy QCD phenomenology, since its contribution is very much suppressed
by the ratio of the corresponding energy scales (squared).

By means of Eqs.(\ref{13}),(\ref{13c}),  in the tree approximation
we obtain for the scalar particle mass
\be\label{h_mass1}
(m_H^0)^2 = (130\pm 15~{\rm GeV})^2\,.
\ee
Here, we have assigned 10\% uncertainty into the ratio light quark condensate and
its constituent mass.

The tentative estimate of the Higgs boson mass given above is rather crude.
In order to improve this value we consider below the contributions
of other condensates at the tree level.
The mass can be also affected by radiative corrections which will be
analyzed elsewhere.
Under the assumption of $\gamma_0$ universality,
the normal ordering of the field operators
$HH = :HH: + \langle HH\rangle$ yields
\be
\label{27a}
\frac{\langle HH\rangle}{{m^2_H}}
=\gamma_0.
\ee
The normal ordering of the vector fields $V_iV_j$
defines the vector field condensates normalized on each degree of freedom
\be\label{27} \langle VV\rangle
={M^2_V}\cdot \gamma_0,~~~V=W^\pm,Z\,,
\ee
calculated in the gauge $V_0=0$. Here, $M_V$ is a corresponding
mass of the vector field. Transverse and longitudinal components are considered
on equal footing  in the reduced phase space
quantization of the massive vector theory~\cite{h-pp}.
As a result, one obtains the upper limit of the vector field condensate contributions
for the mass formula (\ref{h_mass1})
at the tree level for the SM
\be\label{28}
\Delta m_H^2\!=\!\frac{3\lambda^2}{4}\!\langle HH\rangle\!+\frac{3}{8}g^2\!
\left(2\langle WW\rangle+\frac{\langle ZZ\rangle}{\cos^2\theta_W}\!\right),
\ee
where $g$ and $\theta_W$ are the Weinberg coupling constant and the mixing angle.
In Eq.(\ref{28}) the first term is a contribution to the square mass due to
the very scalar field condensate $\langle HH\rangle$.
Taking into account the values of coupling constants, mixing angle, masses,
and condensates, we arrive to the following result
\be
m_H=m^0_H{\left[1+4\frac{\Delta m_H^2}{v^2}\right]}^{1/2}\approx
m^0_H\cdot \left(1+0.02\right),
\ee
where $m^0_H$ is given by Eq.(\ref{h_mass1}).
If there exist additional heavy fields interacting with the SM Higgs boson,
their condensates would contribute to the Higgs boson mass.

In this way, we suggest the condensate mechanism of spontaneous conformal symmetry breaking in the Standard Model.  
We suppose that this mechanism in its origin
is related to the vacuum Casimir energy.
This enables us to avoid the problem of the
regularization of the divergent tadpole loop integral.
The top quark condensate supersedes the phenomenological
negative square mass term in the Higgs potential. The number of free parameters
in the SM lagrangian is reduced, since the tachyon mass term is dropped.  
The condensate mechanism allows to establish relations between condensates and masses including the Higgs boson one.

To get numerical results we use the additional assumption that 
the condensates of all SM fields normalized to their
masses (to the proper power) and degrees of freedom 
represent a universal conformal invariant. 
This allows to to estimate the value of the top quark condensate and
consequently the Higgs boson mass in agreement with recent experimental data.

\subsection*{Acknowledgments}
The authors would like to thank  A.~Efremov and Yu.A.~Budagov
for useful discussions. ABA thanks for support the Dynasty foundation. VNP was supported in part by the Bogoliubov--Infeld
program. AEP and AFZ are grateful to the JINR Directorate for hospitality.

\end{document}